\documentclass[pra,twocolumn,superscriptaddress]{revtex4-2}

\usepackage[pdftex]{graphicx}
\usepackage{dcolumn}
\usepackage{bm}
\usepackage{svg}
\usepackage[utf8]{inputenc}
\usepackage[T1]{fontenc}
\usepackage{mathptmx}
\usepackage{braket}
\usepackage{chemformula}
\usepackage{ulem}
\usepackage{siunitx}
\usepackage{amsmath,amssymb}
\usepackage{xcolor}
\DeclareMathAlphabet{\mathcal}{OMS}{cmsy}{m}{n}
\usepackage[colorlinks=true, allcolors=blue]{hyperref}

\begin{document}

\title{Deployment-ready quantum key distribution over a classical network infrastructure in Padua}

\author{Marco Avesani}
\thanks{These authors contributed equally to this work.}
\affiliation{Dipartimento di Ingegneria dell'Informazione, Universit\`a degli Studi di Padova, via Gradenigo 6B, IT-35131 Padova, Italy}

\author{Giulio Foletto}
\thanks{These authors contributed equally to this work.}
\affiliation{Dipartimento di Ingegneria dell'Informazione, Universit\`a degli Studi di Padova, via Gradenigo 6B, IT-35131 Padova, Italy}

\author{Matteo Padovan}
\thanks{These authors contributed equally to this work.}
\affiliation{Dipartimento di Ingegneria dell'Informazione, Universit\`a degli Studi di Padova, via Gradenigo 6B, IT-35131 Padova, Italy}

\author{Luca Calderaro}
\thanks{Currently with ThinkQuantum S.r.l.}
\affiliation{Dipartimento di Ingegneria dell'Informazione, Universit\`a degli Studi di Padova, via Gradenigo 6B, IT-35131 Padova, Italy}

\author{Costantino Agnesi}
\affiliation{Dipartimento di Ingegneria dell'Informazione, Universit\`a degli Studi di Padova, via Gradenigo 6B, IT-35131 Padova, Italy}

\author{Elisa Bazzani}
\affiliation{Dipartimento di Ingegneria dell'Informazione, Universit\`a degli Studi di Padova, via Gradenigo 6B, IT-35131 Padova, Italy}

\author{Federico Berra}
\affiliation{Dipartimento di Ingegneria dell'Informazione, Universit\`a degli Studi di Padova, via Gradenigo 6B, IT-35131 Padova, Italy}

\author{Tommaso Bertapelle}
\affiliation{Dipartimento di Ingegneria dell'Informazione, Universit\`a degli Studi di Padova, via Gradenigo 6B, IT-35131 Padova, Italy}

\author{Francesco Picciariello}
\affiliation{Dipartimento di Ingegneria dell'Informazione, Universit\`a degli Studi di Padova, via Gradenigo 6B, IT-35131 Padova, Italy}

\author{Francesco B.L. Santagiustina}
\affiliation{Dipartimento di Ingegneria dell'Informazione, Universit\`a degli Studi di Padova, via Gradenigo 6B, IT-35131 Padova, Italy}

\author{Davide Scalcon}
\affiliation{Dipartimento di Ingegneria dell'Informazione, Universit\`a degli Studi di Padova, via Gradenigo 6B, IT-35131 Padova, Italy}

\author{Alessia Scriminich}
\affiliation{Dipartimento di Ingegneria dell'Informazione, Universit\`a degli Studi di Padova, via Gradenigo 6B, IT-35131 Padova, Italy}

\author{Andrea Stanco}
\affiliation{Dipartimento di Ingegneria dell'Informazione, Universit\`a degli Studi di Padova, via Gradenigo 6B, IT-35131 Padova, Italy}

\author{Francesco Vedovato}
\affiliation{Dipartimento di Ingegneria dell'Informazione, Universit\`a degli Studi di Padova, via Gradenigo 6B, IT-35131 Padova, Italy}

\author{Giuseppe Vallone}
\affiliation{Dipartimento di Ingegneria dell'Informazione, Universit\`a degli Studi di Padova, via Gradenigo 6B, IT-35131 Padova, Italy}
\affiliation{Dipartimento di Fisica e Astronomia, Universit\`a degli Studi di Padova, via Marzolo 8, IT-35131 Padova, Italy}
\affiliation{Padua Quantum Technologies Research Center, Universit\`a degli Studi di Padova, via Gradenigo 6B, IT-35131 Padova, Italy}

\author{Paolo Villoresi}
\affiliation{Dipartimento di Ingegneria dell'Informazione, Universit\`a degli Studi di Padova, via Gradenigo 6B, IT-35131 Padova, Italy}
\affiliation{Padua Quantum Technologies Research Center, Universit\`a degli Studi di Padova, via Gradenigo 6B, IT-35131 Padova, Italy}

\begin{abstract}
Current technological progress is driving Quantum Key Distribution towards a commercial and worldwide scale expansion. Its capability to deliver unconditionally secure communication will be a fundamental feature in the next generations of telecommunication networks. Nevertheless, demonstrations of QKD implementation in a real operating scenario and their coexistence with the classical telecom infrastructure are of fundamental importance for reliable exploitation. Here we present a Quantum Key Distribution application implemented over a classical fiber-based infrastructure. By exploiting just a single fiber cable for both the quantum and the classical channel and by using a simplified receiver scheme with just one single-photon detector, we demonstrate the feasibility of low-cost and ready-to-use Quantum Key Distribution systems compatible with standard classical infrastructure.
\end{abstract}

\maketitle

\section{Introduction}
Quantum Key Distribution (QKD) has become one of the main actors of the second quantum revolution and its growth towards commercialization is rapidly bringing quantum technologies into everyday life. With the aim to improve the current security protocols with information-theoretical security guaranteed by physical laws \cite{Bennett2014_BB84,Scarani2008,Pirandola2019rev,Sidhu2021}, QKD is the focus of many international projects whose purpose is the development of a quantum network \cite{Riedel_2019, Raymer_2019,Mehic2020}. To achieve this goal, the possibility to exploit the already existing telecommunication infrastructure is a clear advantage.
Consequently, in recent years, the research in QKD has been focusing on reliability, robustness, cost-effectiveness, compactness, and ease of deployment \cite{Duligall_2006, Sibson:17, Xia:2019, Agnesi:20, Avesani:21}.

\begin{figure*}[t]
\centering
\includegraphics[width = 0.9\linewidth]{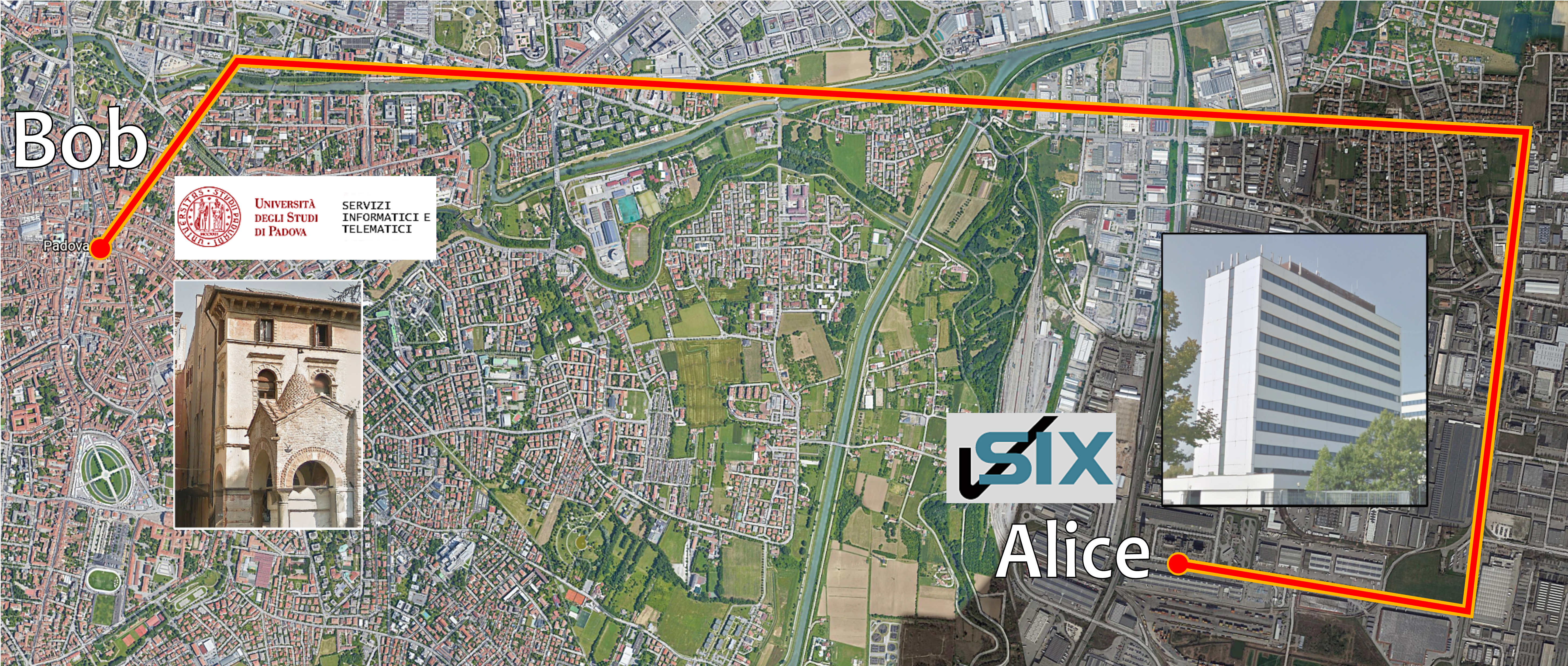}
\caption{\label{fig:map} Aerial view of the fiber link used in the QKD field trial performed in Padua, Italy. [\copyright2021 Google]. The link connects the QKD transmitter, placed at the VSIX center, with the QKD receiver, placed at the ICT Center of UniPD, over deployed metropolitan fibers, for a total of 13~km. }
\end{figure*}

In this work, we present a QKD field trial mutiplexed with classical communication over a single 13km-long telecom dark fiber, which connects the ICT center of the University of Padova, with an Internet Exchange Point: the \textit{Centro di Ateneo per la Connettività e i Servizi al Territorio} VSIX (University Center for Connectivity and Services for the Territory). The two nodes are located respectively in the city-center and industrial area of Padova (Fig. \ref{fig:map}).

In this experiment we implement the efficient three-state one-decoy BB84 protocol \cite{Grunenfelder2018} which encodes the information in the polarization of weak coherent pulses. The transmitter relies on a POGNAC \cite{Agnesi2019} encoder, a simple yet stable Sagnac-based polarization modulator characterized by a low intrinsic QBER. Moreover, the temporal synchronization employed the Qubit4Sync procedure \cite{Calderaro2020}. Both these features were previously developed and tested \cite{Agnesi:20, Avesani:21} by our research group and already matched the requirements of simplicity and robustness of deployed QKD nodes. Nevertheless, the system here presented provides significant improvements with respect to the recent tests of Ref. \cite{Avesani:21}.

To ease the system deployment, both the transmitter and receiver were built to fit a standard 4U rack case. To maximize compactness, the receiver employs only a single-photon detector, external to the 4U rack, instead of the more common choice of four. Furthermore, both quantum and classical communications were established within the same fiber with wavelength-multiplexing: 1550~nm for the first and 1490~nm for the second.

\begin{figure}
    \centering
    \includegraphics[width = \linewidth]{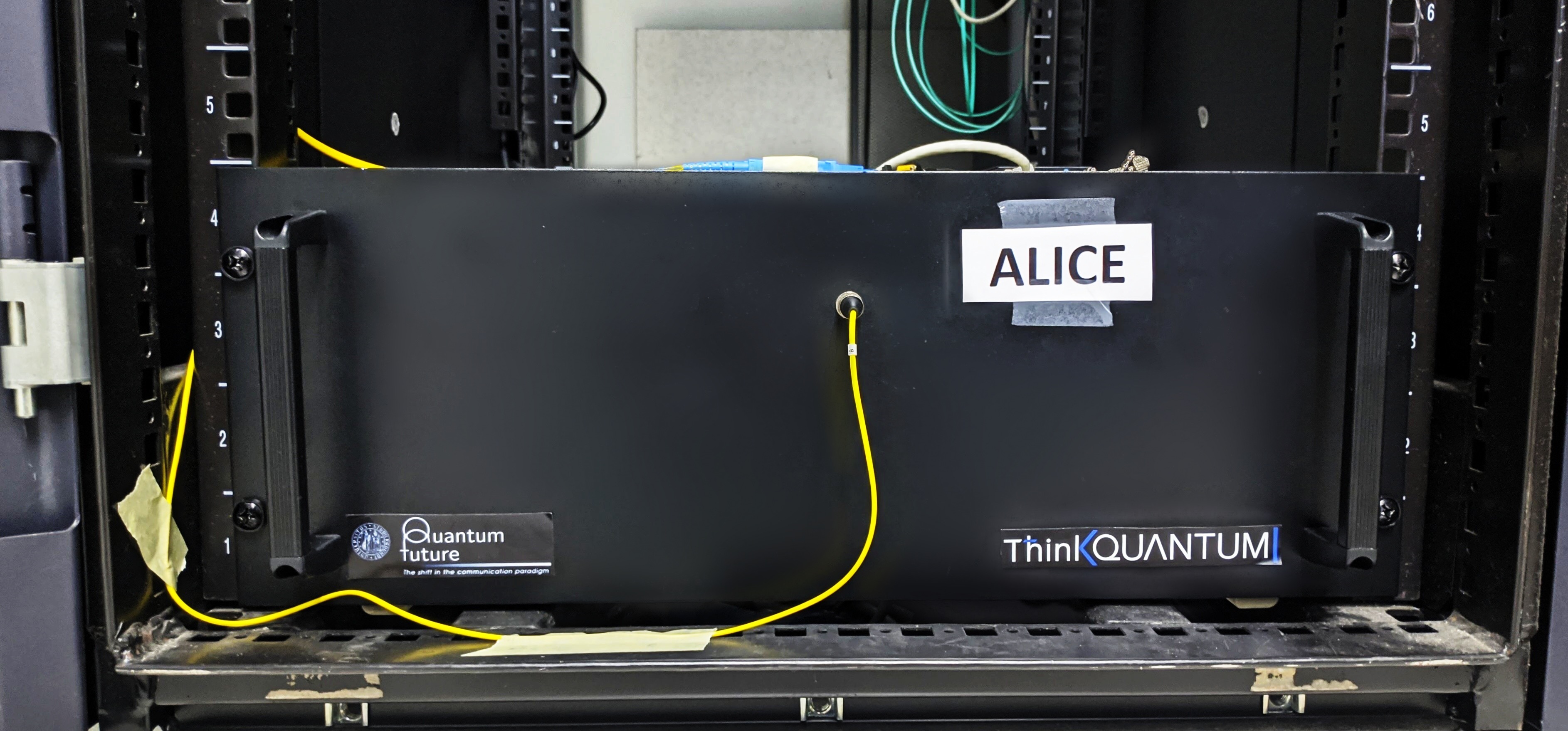}\\
    (a)\\\vspace{0.25cm}
    \includegraphics[width = \linewidth]{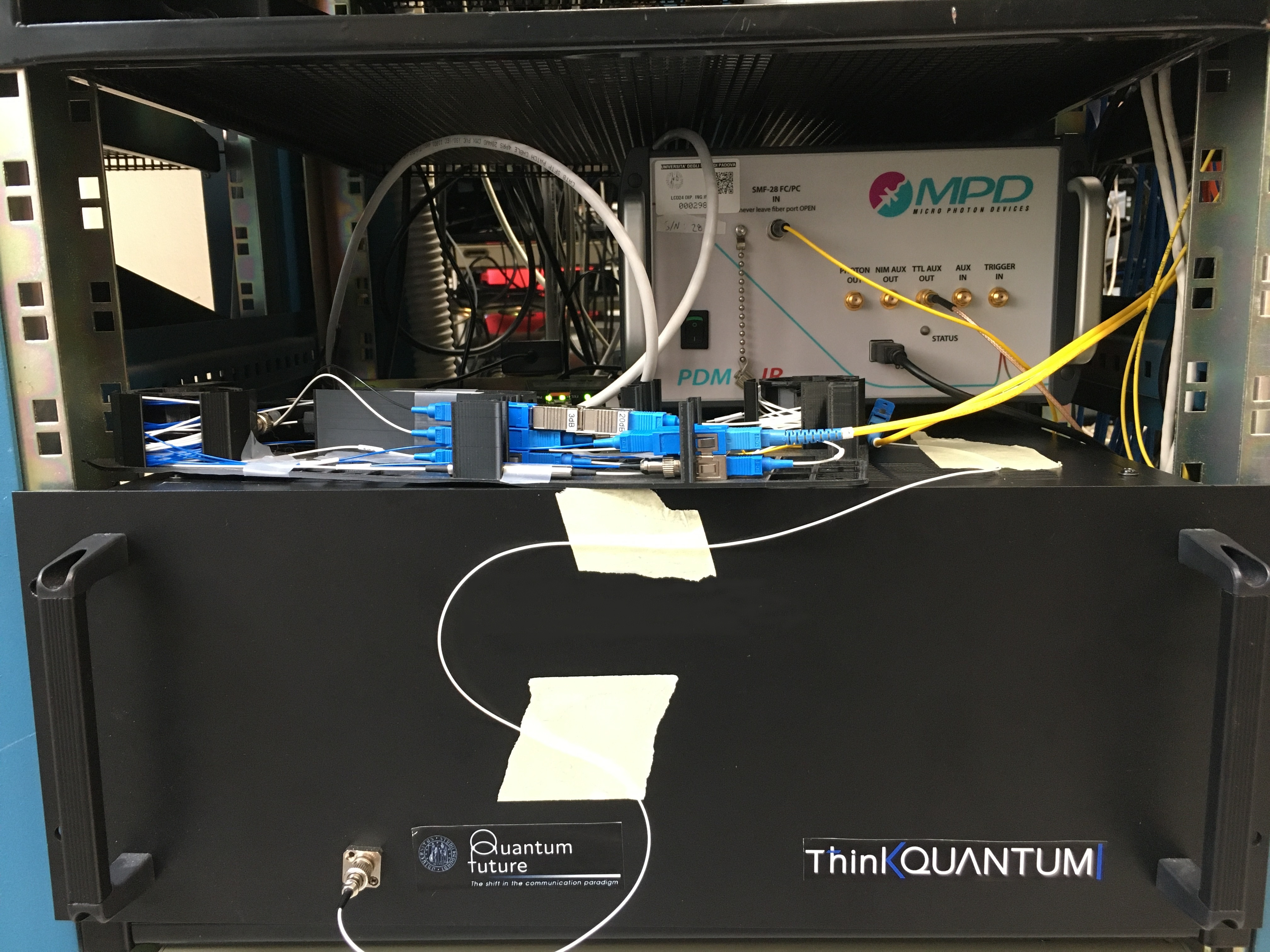}\\
    (b)
    \caption{The modular system implemented for the field-test.}    \label{fig:AliceBobRack}
\end{figure}

These design choices, alongside the criterion of flexibility and modularity pursued when developing the system, enabled us to fully exploit an already stationed telecom dark fiber by deploying on-site solely two boxes in a rack mount (which included the control PCs) plus two modules for the classical communication (see Fig. \ref{fig:AliceBobRack} and \ref{fig:CompleteScheme}). During a $24$-hour long acquisition, we were able to generate $144$ Mb of secret key.

\section{The field trial}
\label{sec:Methods}
\vskip .2 truecm 

\subsection{Infrastructure}
In the field trial, the transmitter (Alice) and the receiver (Bob) are placed respectively at the VSIX facility and the University's ICT center, as shown in Fig.~\ref{fig:map}.
Alice and Bob are connected through a 13km-long optical fiber. The attenuation introduced by the link is approximately 6.7 dB.

\begin{figure*}[t]
\centering
\includegraphics[width =\linewidth]{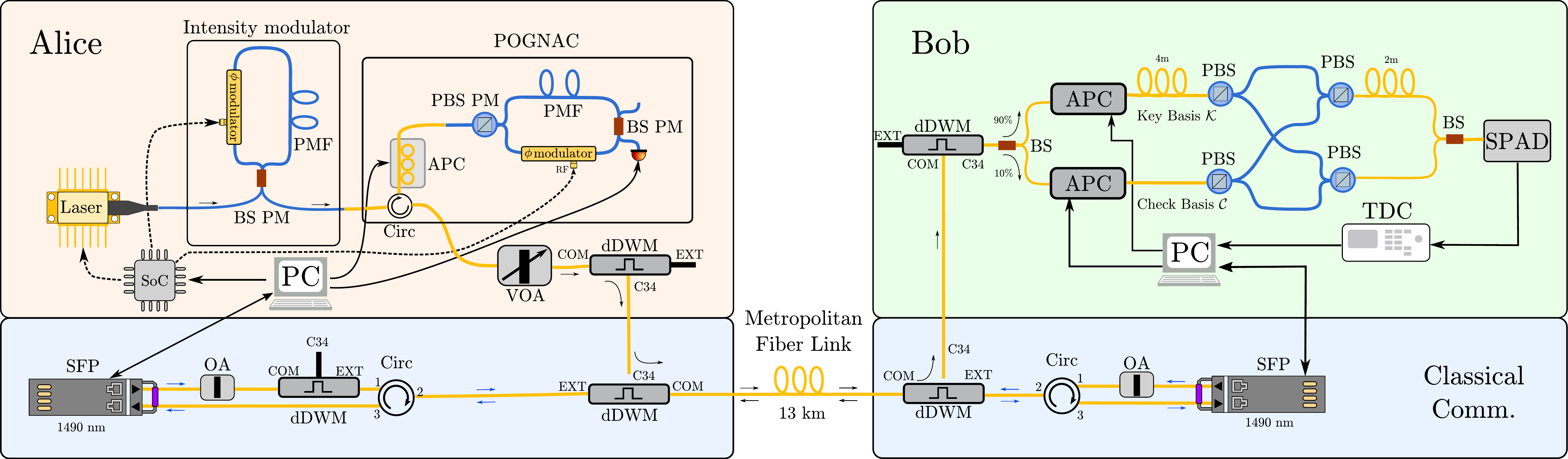}
\caption{Schematic representation of the setup employed in the experiment. In the upper-left and upper-right boxes we represent the setup for the quantum transmitter and quantum receiver, respectively. In the lower boxes we show the components employed for the classical communication and the wavelength multiplexing setup. }
\label{fig:CompleteScheme}
\end{figure*}

Both the classical communication and the transmission of qubits are conveyed through the same fiber.
The classical signal is provided by two standard small form-factor pluggable (SFP) duplex transceivers operating at $1490$ nm in half-duplex mode by means of two optical circulators (Circ).
Moreover, two optical attenuators (OA) lessen the power of both 1490 nm transmitters to the minimum detection threshold ($\approx -31$~dBm at the receiver) to reduce the disturbance on the quantum channel.
To mix and decouple the classical and quantum signals on the same fiber, sequences of dense wavelength division multiplexing (dWDM) filters (bandwidth $\approx$~100~GHz) are adopted.
This scheme is shown in the bottom section of Fig. \ref{fig:CompleteScheme}.

\subsection{Transmitter}

Shown on the left of Fig. \ref{fig:CompleteScheme}, the source is a $1550$~nm gain-switched distributed feedback laser, emitting $270$~ps full-width-at-half-maximum pulses with a repetition rate of $50$~MHz. The optical encoder consists of two main sections: the intensity modulator followed by the polarization encoder.

The intensity modulator is based on a fiber-optic Sagnac loop and includes a 70:30 beamsplitter (BS), a lithium niobate \ch{LiNbO_3} phase modulator, and a $1$m-long delay line~\cite{Roberts2018}. 
This scheme implements the decoy-state protocol by setting two possible mean photon numbers ($\mu \text{ and } \nu$) for the transmitted pulse. These parameters are chosen in such a way that their ratio is $\frac{\mu}{\nu} \approx 3.5$. 
Source and intensity modulators are composed of all fiber polarization-maintaining (PM) devices with the light propagating along to the fibers' slow axis. 

The light polarization is modulated by the POGNAC\cite{Agnesi2019}. Like the intensity modulator, the POGNAC relies on an unbalanced Sagnac interferometer with the beamsplitter replaced by a polarization beamsplitter (PBS). Before entering the PBS, the light passes through an optical circulator and an automatic polarization controller (APC). The APC modifies the polarization to obtain a balanced superposition of vertical and horizontal states with arbitrary relative phase. Hence, the light is equally split into the clockwise and counterclockwise modes of the loop. Thanks to the asymmetry of the interferometer, by properly setting the voltage of the modulator and carefully timing the two pulses, one can generate left 
($\ket{L}$) or right ($\ket{R}$) circular polarization states.
If no phase is applied to any of the two pulses, the resulting state is 
$\ket{D} = \frac{1}{\sqrt{2}} (\ket{H} + \ket{V})$. These three states allow us to implement the 3-state efficient BB84 protocol \cite{Fung2006}, where the keying basis is $\mathcal{K} = \{\ket{L}, \ket{R} \}$ 
and the check basis is $\mathcal{C} = \{\ket{D}, \ket{A} =$ 
$ \frac{1}{\sqrt{2}} (\ket{H} - \ket{V})\}$. To monitor the optical power flowing in the two directions and provide feedback for the APC alignment, a 90:10 BS in the loop directs $90\%$ of the light to a power meter.

The polarization-modulated light travels through the APC and emerges from the exit port of the circulator. Finally, the signal passes through a dWDM filter, is attenuated to the single-photon regime ($\mu = 0.6$, $\nu = 0.17$) by a variable optical attenuator (VOA), and is transmitted over the quantum channel.

The main control core of the transmitter is a System-on-a-Chip (SoC) which comprehends both an FPGA and a CPU integrated on a dedicated board (ZedBoard by Avnet). Thys system is responsible for the synchronized generation of signals to drive the laser and the two electro-optical phase modulators. The signals are produced according to a pseudo-random sequence. The full SoC architecture is described in \cite{Stanco2021}.
The post-processing, communication, and autonomous control of the entire system are performed on PCs by a custom-made software.

\subsection{Receiver}

Shown on the right of Fig. \ref{fig:CompleteScheme}, Bob measures the incoming photons by randomly choosing between the $\mathcal{K}$ and $\mathcal{C}$ bases with equal probability.
Subsequently, an APC and a PBS perform the projective measurement for each basis. Before each run of the QKD protocol, the APCs are aligned by sending a public state sequence and monitoring the resulting QBER. Furthermore, during the key exchange, Alice interleaves four pre-shared and publicly known bits every $36$ private key bits. Bob recognizes and uses them to keep his measurement bases aligned with the APC. Nevertheless, the procedure reduces the achievable generation rate by $10\%$. 

To improve compactness and significantly reduce cost, we adopted a time-multiplexing scheme.
This allowed us to perform the measurements on two bases using only one InGaAs/InP single-photon avalanche diode (SPAD). We used the PDM-IR from Micro Photon Devices S.r.l., which provides $15\%$ quantum efficiency. However, this simplification introduced an additional attenuation of $3$ dB. This time multiplexing feature works by differently delaying the pulses exiting the PBSs and by routing them towards the detector with two additional PBSs and a BS. The time tags corresponding to the photons arrivals are recorded by a time-to-digital converter (quTAU from qutools GmbH) and transmitted to Bob's PC for data processing. The time synchronization is provided by the Qubit4Sync algorithm\cite{Calderaro2020} which reconstructs, according to the detection events, the clock period without any external reference signal.

\section{Results}
\label{sec:Results}

\begin{figure}
    \centering
    \includegraphics[width = \linewidth]{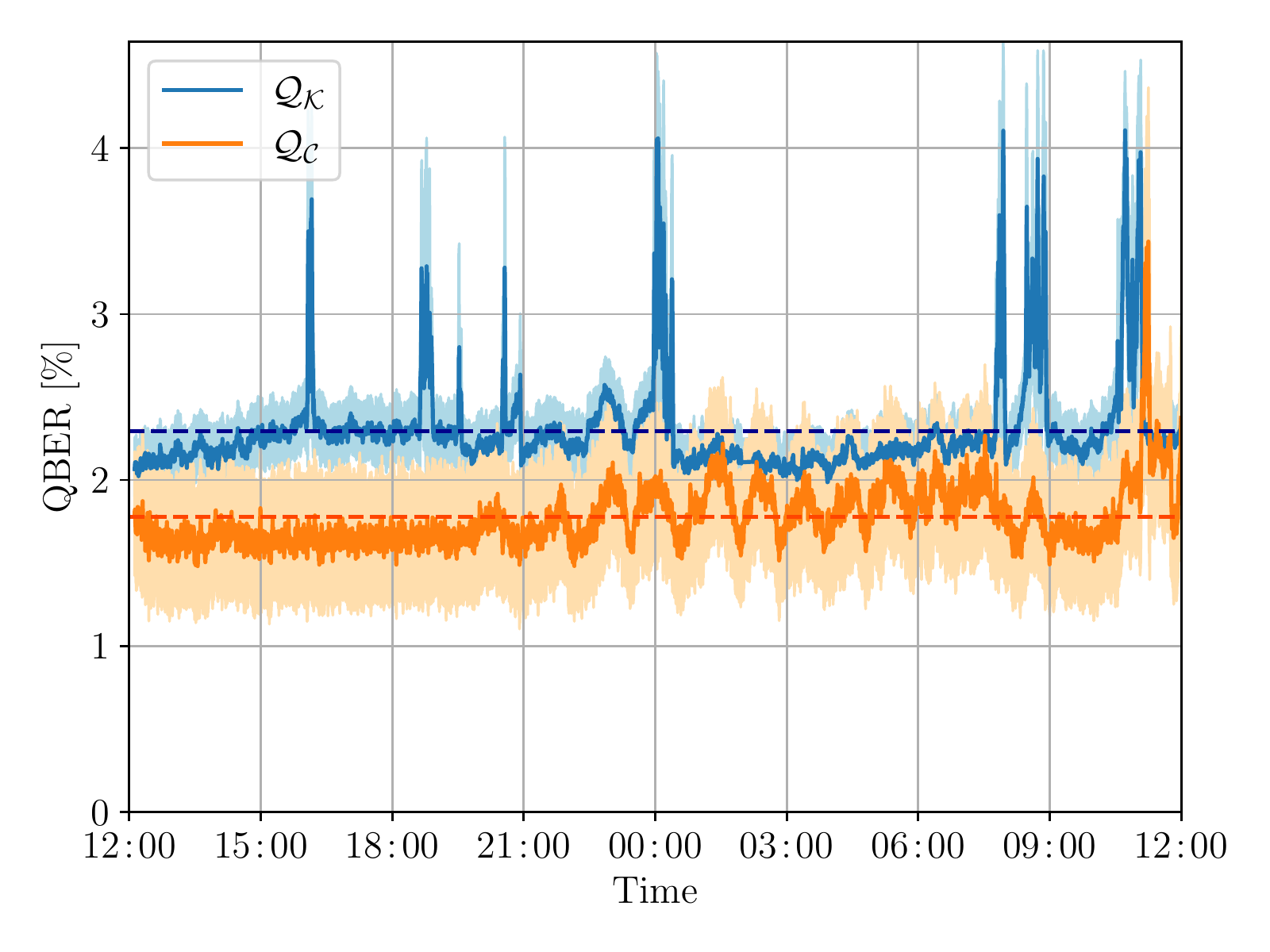}
    \caption{Behaviour of the QBERs in the $\mathcal{K}$ and $\mathcal{C}$ bases during the $24$-hour long acquisition. They are measured every second and a rolling mean ($60$ s window) is plotted. Colored bands represent $\pm 1$ standard deviations, calculated on the rolling window. Dashed lines are the overall averages $2.3\%$ and $1.8\%$.}
    \label{fig:QBER}
\end{figure}

In a $24$-hour long acquisition, we produced $1.94$ Gb of raw key material.
The average quantum bit error rates (QBERs) in the key ($\mathcal{Q_\mathcal{K}}$) and check ($\mathcal{Q_\mathcal{C}}$) bases were $2.3\%$ and $1.8\%$ respectively.
The error correction procedure operates over small key blocks, corresponding to about $1$ s of acquisition.
Any block whose QBER in at least one basis is above $5\%$ is discarded to improve the average QBER of the key.
In Fig. \ref{fig:QBER}, we show the QBERs behavior during the acquisition.
The fluctuations are mostly due to minor temperature changes, but whenever the QBER grows too large (specifically, above $2.5\%$), the automatic polarization alignment mechanism provides to reduce it.
By comparing these results with a previous run in which the classical communication was deactivated, we can infer that the background noise caused by the wavelength multiplexing contributes to about $0.6\%$ of the QBER.

The post-processing, based on a modified version of the AIT QKD R10 software suite, by the AIT Austrian Institute of Technology GmbH \cite{AIT}, ran in parallel with the quantum transmission, starting only after a small delay.
The authentication of all its messages began from a pre-shared random sequence, later replaced by small portions of the keys produced by QKD as soon as they were available.

The production of the secret keys is based on the finite-size analysis of Ref. \cite{Rusca2018}, which is applied to $5\cdot 10^6$-bit long blocks of sifted key in the $\mathcal{K}$ basis.
Each block is shortened in privacy amplification to a secret key length (SKL) of
\begin{equation}
 \mathrm{SKL} = s_{\mathcal{K},0} + s_{\mathcal{K},1}(1 - h(\phi_\mathcal{K})) - \lambda_{\rm EC} -\lambda_{\rm c} - \lambda_{\rm sec} .
\end{equation}
Terms $s_{\mathcal{K},0}$ and $s_{\mathcal{K},1}$ are the lower bounds on the number of vacuum and single-photon detection events in the key basis, $\phi_\mathcal{K}$ is the upper bound on the phase error rate corresponding to single-photon pulses, $h(\cdot)$ is the binary entropy, $\lambda_{\rm EC}$ and $\lambda_{\rm c}$ are the number of bits published during the error correction and confirmation of correctness steps, and finally $\lambda_{\rm sec}=6 \log_2(\frac{19}{\epsilon_{\rm sec}})$, where $\epsilon_{\rm sec}=10^{-10}$ is the security parameter associated with the secrecy analysis.

\begin{figure}
    \centering
    \includegraphics[width = \linewidth]{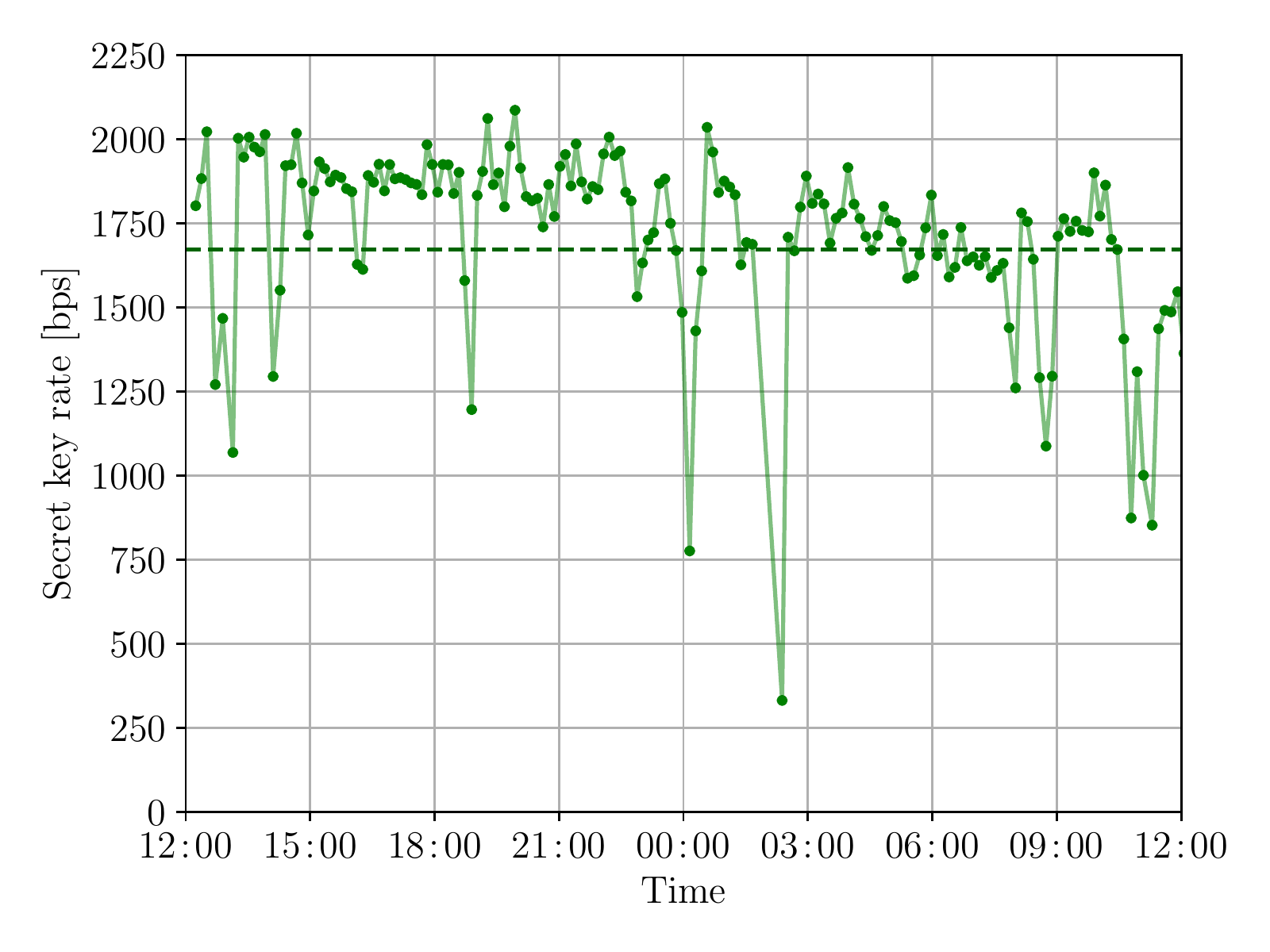}
    \caption{Behaviour of the secret key rate during the $24$-hour long acquisition. The dashed line represents the average: $1.7$ kbps.}
    \label{fig:SKR}
\end{figure}

In Fig. \ref{fig:SKR}, we plot the behavior of the secret key rate (SKR), which is the ratio between the SKL of each block and the time needed to produce it.
The temporary drops can be correlated with the increases of the QBER in Fig. \ref{fig:QBER}.
On average, the SKR was $1.7$ kbps.
This rate was sufficient for the applications required by this system, whose main target was portability, low-cost, and low-complexity. However, thanks to the modularity of the implementation described in this work, greater rates are achievable if a source with higher repetition rates and more detectors are employed.

\section{Conclusion}
\label{sec:Conclusion}
The QKD system here presented has several features that make it simple and robust.
In addition to a stable and accurate polarization encoder and a software-based synchronization approach, it uses wavelength-multiplexing to combine quantum and classical communication in the same optical fiber and time-multiplexing to use only one single-photon detector.

The field trial over an already existing 13km-long fiber is proof of the efficacy of these solutions and their readiness for real applications.
It represents an important step towards the deployment of QKD networks in an urban setting.

\begin{acknowledgments}
    Part of this work was supported by European Union’s Horizon 2020 research and innovation programme, project OpenQKD (grant agreement No 857156), by Agenzia Spaziale Italiana, project {\it Q-SecGroundSpace} (Accordo n. 2018-14-HH.0, CUP: E16J16001490001) and by Ministero dell'Istruzione, dell'Università e della Ricerca (MIUR) (Italian Ministry of Education, University and Research) under the initiative ``Departments of Excellence'' (Law No. 232/2016).
    
    We thank Eng. L. Finotti, Eng. A. Celin and Eng. C. Marangon of VSIX together with Eng. G. Paolucci and Eng. F. Ballarin of ASIT – UniPD for the logistics support.  
    We also thank Dr. C. Pacher and Dr. O. Maurhart and the AIT Austrian Institute of Technology GmbH for providing the foundation for the post-processing software.
    CloudVeneto is acknowledged for the cloud resources.
\end{acknowledgments}

\bibliography{references.bib}

\end{document}